# The modified Klein Gordon equation for neolithic population migration

M. Pelc[1] J. Marciak – Kozlowska [2] M. Kozlowski [3]

[1] Postgraduate student, Physics Institute, Maria Curie-Sklodowska University, Lublin, Plac Marii Curie Sklodowskiej, Lublin Poland
[2] Institute of Electron Technology, Al. Lotnikow 32/46, 02-668 Warsaw, Poland
[3] Corresponding author, e-mail:miroslawkozlowski@aster.pl

Abstract

In this paper the model for the neolithic migration in Europe is developed. The new migration equation, the modified Klein Gordon equation is formulated and solved. It is shown that the migration process can be described as the hyperbolic diffusion with constant speed. In comparison to the existing models based on the generalization of the Fisher approach the present model describes the migration as the transport process with memory and offers the possibility to recover the initial state of migration which is the wave motion with finite velocity.

## 1. Introduction

The origins of European agriculture are normally sought in the Near East. The earliest indicators of agriculture in the form of cultivation of cereals and pulses are rearing of animals come from Zagros foothills. Their age, 11700 – 8400 BC corresponds to the cool dry climatic period followed by a rapid increase in rainfall at the beginning of the Holocene (~ 10000 BC).

During the early stages of agricultural development (Preceramic Neolithic 9800 – 7500 BC) the rapid increase in the number of sites is noticeable in both the foothills and the surrounding plains accompanied by the appearance of large settlements with complicated masonry structures and fortifications (e.g. Jericho).

At a later stage the core area of early agricultural settlements shifts to the north to the eastern highlands and inner depressions of Asia Minor. The most outstanding case of early agricultural development in this area is Catal Hüyük, a Neolithic town on the Konya Plain (6500 – 5700 BC).

The earliest sites with developed agricultural economies in Europe, dated 6400 – 6000 BC are found in the intermontane depressions of Greece (Thessaly, Beotia and Pelloponesse). Genetic features of the cultigens and the general character of the material culture leave no doubt to their Near – Eastern origins. Significantly, the early Neolithic sites in the Marmara Sea are of a more recent age (6100 – 5600 BC), being culturally distinct from the Early Neolithic in Greece. This implies that the Neolithic communities could penetrate the Balkan Peninsula from Western Asia by means of navigation.

The Neolithic spread further, plausibly via the Strouma axis in the northeast and the Vardar – Morava axis in the north. The ensuing development saw a rapid growth of Neolithic settlements in the depressions of northern Thrace the Lower and Middle Danube catchment basin (5900 – 5500 BC).

The next stage in the Neolithic development saw the emergence of new sites on the Tisza Plain in 5600 – 5500 BC [1].

The new sites later spread over the vast areas of the loess plains of Central Europe, mostly along the Danube, Rhine – Mainz and Vistula axes. This spread occurred within the range of 5600 – 4800 BC within the most probable age of 5154 ± 62 BC [2].

Judging from the number of sites the population in the Near East started increasing ~ 15000 BC. *Ammerman and Cavalli – Sforza* [3] focused on measuring the rate of spread of early farming in Europe and derived the rate of spread $v \sim 1$ km/year on average in Europe.

The Danube and Rhine valleys the propagation paths had an increased propagation speed as did the Mediterranean west [4]. The speeds of propagation of the wave front $v$ in these areas are as follows:

$v$ = 1 km/year on average in Europe

$v$ = 4 – 6 km/year for the Danube – Rhine valleys

$v$ = 10 km/year for Mediterranean regions

Interpretations of these observations are usually based on the reaction – diffusion equation of population dynamics [5,6]. Fort and Méndez [6,7] discussed the front propagation rate resulting from the generalization of Fisher model, but their results are restricted to the homogenous system.

The aim of this work is to formulate and develop a model for the spread of incipient farming in Europe taking into account the environmental influences on the migration processes.

## 2. The model

Fick diffusion equation is a special case of the parabolic transport equation in which speed of perturbation propagation is infinite. Parabolic transport equation has been applied to the spread of advantageous genes [8] dispersion of biological population [9], epidemic models [10].

However if perturbation propagates at finite speed, Fick law does not hold [11]. This unphysical feature can be avoided by making use of hyperbolic transport equation [11]. The hyperbolic propagation equation, HPE has been very recently applied to the spread of epidemics [12], forest fire models [13] and chemical system [14].

An interesting application of the Fick law to the migration in neolithic Europe was presented in [15].

Such a model as was presented in [15] provides a consistent explanation for the origin of Indo – European languages [16] and also finds remarkable support from the observed gene frequencies [17].

In this paper we develop the model for the population migration following the method presented in [11].

Let $n(x,t)$ stands for population density (measured in number of families per square kilometer) where $x$ is Cartesian coordinate and $t$ is time. We assume that a well defined time scale, $\tau$ between two successive migration steps exists. The migrated population interacts with environment. This interaction we will model by potential $V$. in that case the hyperbolic migration equation can be written as [11]:

$$\tau \frac{\partial^2 p}{\partial t^2} + \frac{\partial p}{\partial t} + \frac{V p_k}{D E_k} p = D \frac{\partial^2 p}{\partial x^2} + G(x,t) \tag{1}$$

In equation (1) $D$ is the migration diffusion coefficient, $p_k$ and $E_k$ denote the momentum and kinetic energy of migration and $G(x,t)$ is the population growth.

Equation (1) is the Heaviside equation for population density [4]. It is the hyperbolic equation which describes the migration with memory. Memory term $\tau \partial^2 p / \partial t^2$ changes the type of the equation. For $\tau = 0$ equation [1] is the parabolic Fick equation for population migration.

We seek solution of the Eq.(1) in the form

$$p(x,t) = e^{-t/2\tau} u(x,t) \tag{2}$$

After substituting of Eq. (2) to Eq.(1) one obtains

$$\frac{1}{v^2}\frac{\partial^2 u}{\partial t^2} - \frac{\partial^2 u}{\partial x^2} + u(x,t)\left[-\frac{1}{4\tau D} + \frac{V p_k^2}{D^2 E_k}\right] = \frac{G(x,t)}{D} e^{t/2\tau}. \tag{3}$$

For $V < 0$, i.e. for attractive potential equation (3) is the modified Klein – Gordon equation, i.e. so called “telegrapher equation”. For $V > 0$ we have two possibilities:

$$\tau > \frac{E_k D}{4 V p_k^2} \tag{4}$$

and

$$\tau < \frac{E_k D}{4 V p_k^2} \tag{5}$$

In the case described by the inequality (4) equation (3) is the Klein – Gordon equation and for (5) Eq. (3) is the telegrapher equation.

In the following we will consider the $V < 0$, i.e. attractive potential; in that case equation (3) can be written as

$$\frac{1}{v^2}\frac{\partial^2 u}{\partial t^2} - \frac{\partial^2 u}{\partial x^2} - q u(x,t) = F(x,t). \tag{6}$$

$$q = \frac{1}{4\tau D} + \frac{|V| p_k^2}{D^2 E_k}, \qquad q > 0 \tag{7}$$

and

$$F(x,t) = \frac{G(x,t)e^{t/2\tau}}{D} \tag{8}$$

Let us consider the initial condition

$$u(x,t) = f(x), \qquad u_t(x,0) = g(x) \\ -\infty < x < \infty \tag{9}$$

for the equation (6). Then the general solution of Eq. (6) can be written as [18]:

$$\begin{aligned} u(x,t) &= \frac{f(x-vt)+f(x+vt)}{2} \\ &+ \frac{1}{2v}\int_{x-vt}^{x+vt} g(\varsigma) I_0\left[\left(\tfrac{q}{v}\right)^{1/2}\sqrt{v^2t^2-(x-\varsigma)^2}\right]d\varsigma \\ &- \frac{(qv)^{1/2}}{2} t \int_{x-vt}^{x+vt} f(\varsigma)\frac{I_1\left[\left(\tfrac{q}{v}\right)^{1/2}\sqrt{v^2t^2-(x-\varsigma)^2}\right]}{\sqrt{v^2t^2-(x-\varsigma)^2}}d\varsigma \\ &+ \frac{1}{2v}\int_0^t \int_{x-v(t-\eta)}^{x+v(t-\eta)} F(\varsigma,\eta) v I_0\left[\left(\tfrac{q}{v}\right)^{1/2}\sqrt{v^2(t-\eta)^2-(x-\varsigma)^2}\right]d\varsigma d\eta. \end{aligned} \tag{10}$$

Considering Eq. (10) and Eq. (2) the solution of Eq. (1) can be written as:

$$p(x,t) = e^{-t/2\tau}u(x,t) \tag{11}$$

where the $u(x,t)$ is described by formula (10).

As can be seen from formula (11) the general solution of Eq. (1) is the sum of the wave motion described by the term

$$\frac{f(x-vt)+f(x+vt)}{2}$$

and the lag term.

In conclusion one can say that the migration can be described as the damped wave motion which for $t \to \infty$ is going to diffusion of the population.